# On the Numerical Stationary Distribution of Overdamped Langevin Equation in Harmonic System

De-Zhang Li[1] and Xiao-Bao Yang[1*]

[1] Department of Physics, South China University of Technology, Guangzhou 510640, China.

* Corresponding authors. Correspondence and requests for materials should be addressed to X.-B. Y. (email: scxbyang@scut.edu.cn).

## Abstract

Efficient numerical algorithm for stochastic differential equation has been an important object in the research of statistical physics and mathematics for a long time. In this paper we study the highly accurate numerical algorithm of the overdamped Langevin equation. In particular, our interest is the behaviour of the numerical schemes for solving the overdamped Langevin equation in the harmonic system. Three algorithms are obtained for overdamped Langevin equation, from the large friction limit of the schemes for underdamped Langevin dynamics. We derive the explicit expression of the stationary distribution of each algorithm by analysing the discrete time trajectory, for both one-dimensional and multi-dimensional cases. The accuracy of the stationary distribution of each algorithm is illustrated by comparing to the exact Boltzmann distribution. Our results demonstrate that, the "BAOA-limit" algorithm generates the exact distribution for the harmonic system in the canonical ensemble, within the stable regime of the time interval. The other algorithms do not produce the exact distribution of the harmonic system.

# 1. Introduction

As the first published stochastic differential equation, Langevin equation[1, 2] plays an important role in statistical mechanics and mathematical physics. It offers a heuristic mathematical description of the Brownian motion[3-5], and becomes a widely used tool in the fields of natural science, mathematics and social science[6]. Specifically, the Boltzmann distribution in the canonical ensemble can be obtained from the stationary state distribution of Langevin equation[7]. Hence, Langevin equation is an efficient approach to the problem of canonical sampling in statistical mechanics.

In this paper, we focus on a certain type of Langevin equation, called overdamped Langevin equation. It can be seen as the overdamped limit of the usual underdamped Langevin equation[8]. For the system with potential energy $U$ and mass matrix $\mathbf{M}$, the overdamped Langevin equation takes the form

$$\frac{d\mathbf{x}}{dt} = -\frac{1}{\gamma}\mathbf{M}^{-1}\nabla U(\mathbf{x}) + \sqrt{\frac{2}{\beta\gamma}}\mathbf{M}^{-1/2}\boldsymbol{\eta}(t) \ . \tag{1}$$

Here $\gamma$ is the friction coefficient, $\beta = 1/k_B T$ with the Boltzmann constant $k_B$ and the temperature $T$, and $\boldsymbol{\eta}(t)$ is the white noise random vector associated with the Wiener process satisfying

$$\langle \eta_i(t) \rangle = 0, \ \langle \eta_i(t)\eta_j(t') \rangle = \delta_{ij}\delta(t-t') \tag{2}$$

or equivalently

$$\langle \boldsymbol{\eta}(t) \rangle = 0, \ \langle \boldsymbol{\eta}(t)\boldsymbol{\eta}^T(t') \rangle = \delta(t-t')\mathbf{1} \ . \tag{3}$$

$\mathbf{1}$ denotes the identity matrix hereinafter. Consider the probability density function of the system $\rho(\mathbf{x},t)$, associated with Eq. (1). The time-dependent probability density $\rho(\mathbf{x},t)$ of the system evolves according to the well-known Fokker-Planck equation [6, 7, 9-12]

$$\frac{\partial}{\partial t}\rho(\mathbf{x},t) = \frac{\partial}{\partial \mathbf{x}} \cdot \left[ \frac{1}{\gamma} \mathbf{M}^{-1} \frac{\partial U}{\partial \mathbf{x}} \rho(\mathbf{x},t) + \frac{1}{\beta\gamma} \mathbf{M}^{-1} \frac{\partial \rho(\mathbf{x},t)}{\partial \mathbf{x}} \right]. \tag{4}$$

The relevant time evolution operator is then

$$\mathcal{L}(\sim) = \frac{\partial}{\partial \mathbf{x}} \cdot \left[ \frac{1}{\gamma} \mathbf{M}^{-1} \frac{\partial U}{\partial \mathbf{x}} (\sim) + \frac{1}{\beta\gamma} \mathbf{M}^{-1} \frac{\partial (\sim)}{\partial \mathbf{x}} \right]. \tag{5}$$

$\rho(\mathbf{x},t)$ evolves from an initial distribution (e.g., $\rho(\mathbf{x},0) = \delta(\mathbf{x}-\mathbf{x}_0)$) to a stationary state $\rho(\mathbf{x},\infty)$, which satisfies $\mathcal{L}\rho(\mathbf{x},\infty) = 0$. The stationary distribution is usually called the invariant distribution in the mathematical literatures[13, 14]. We denote the stationary distribution as $\rho_{st}(\mathbf{x})$ in this paper. In our case, $\rho_{st}(\mathbf{x})$ is simply the Boltzmann distribution

$$\rho_{st}(\mathbf{x}) = \frac{1}{N} \exp\left[-\beta U(\mathbf{x})\right] \tag{6}$$

where $N$ is the normalization constant. Then one can see that the term $\sqrt{\frac{2}{\beta\gamma}} \mathbf{M}^{-1/2}$ associated with $\boldsymbol{\eta}(t)$ in Eq. (1) fulfils the fluctuation-dissipation relation, which guarantees that the overdamped Langevin equation generates the Boltzmann distribution in the stationary state. Therefore, overdamped Langevin equation can be an effective tool to sample the equilibrium canonical ensemble.

The aim of this work is to study the stationary distributions of the algorithms for numerically solving the overdamped Langevin equation. Various numerical algorithms for overdamped Langevin dynamics (also called Brownian dynamics) have been proposed in previous works [8, 13-17]. Systematic errors arise from the finite time interval $\Delta t$ of the numerical algorithms. Our interest is the exact solution of the stationary distribution with finite $\Delta t$, in the harmonic system. In Section 2, we introduce three algorithms based on the splitting method of the integrators for underdamped Langevin equation. In Section 3, the exact stationary distribution of each algorithm in one-dimensional harmonic system is derived. We employ the stochastic analysis of the discrete time trajectory in the configurational space. The extension of the results to the multi-dimensional harmonic system is done in Section 4, by making use of the normal mode coordinate transformation. The accuracy of the stationary distribution generated from each algorithm is illustrated. Conclusions are outlined in Section 5.

## 2. Numerical algorithms

First we recall the splitting method to construct the integrator of underdamped Langevin equation by Leimkuhler et al.[13],

$$\begin{bmatrix} \dfrac{d\mathbf{x}}{dt} \\ \dfrac{d\mathbf{p}}{dt} \end{bmatrix} = \underbrace{\begin{bmatrix} \mathbf{M}^{-1}\mathbf{p} \\ 0 \end{bmatrix}}_{A} + \underbrace{\begin{bmatrix} 0 \\ -\nabla U(\mathbf{x}) \end{bmatrix}}_{B} + \underbrace{\begin{bmatrix} 0 \\ -\tilde{\gamma}\mathbf{p} + \sqrt{\dfrac{2\tilde{\gamma}}{\beta}}\mathbf{M}^{1/2}\mathbf{\eta}(t) \end{bmatrix}}_{O} \qquad (7)$$

where $\tilde{\gamma}$ is the friction coefficient, the random vector $\mathbf{\eta}(t)$ satisfies the conditions in Eq. (3) and "O" represents the Ornstein-Uhlenbeck [18, 19] part. Each of the three parts of underdamped Langevin equation can be solved "exactly", and the integrator can be obtained from a certain composition of these three parts. As an example, BAOAB denotes the integrator of the order of composition $B\left(\dfrac{1}{2}\delta t\right) A\left(\dfrac{1}{2}\delta t\right) O(\delta t) A\left(\dfrac{1}{2}\delta t\right) B\left(\dfrac{1}{2}\delta t\right)$, where we use $\delta t$ as the time interval for the underdamped Langevin dynamics. It is straightforward to verify that $B\left(\dfrac{1}{2}\delta t\right) A(\delta t) B\left(\dfrac{1}{2}\delta t\right)$ is the famous velocity-Verlet algorithm[20] for integrating the Hamiltonian equation. Alternative splitting methods which divide the underdamped Langevin equation into two parts have also been discussed[21, 22]. Among the integrators constructed from the splitting methods, BAOAB scheme has been shown to produce the most accurate stationary distribution in the configurational space[13, 21-28]. Recently, a scheme named leap-flog "middle" is studied[29, 30]. This scheme is equivalent to BAOA in the stationary state and leads to the same configurational distribution with that of BAOAB. The numerical algorithms of overdamped

Langevin equation considered in this paper can be obtained from the underdamped Langevin integrators constructed from Eq. (7).

The BAOA scheme of underdamped Langevin dynamics is expressed as

$$\begin{aligned}
\mathbf{p}_{n,1} &= \mathbf{p}_n - \delta t \nabla U(\mathbf{x}_n) \\
\mathbf{x}_{n,1} &= \mathbf{x}_n + \frac{\delta t}{2}\mathbf{M}^{-1}\mathbf{p}_{n,1} \\
\mathbf{p}_{n+1} &= e^{-\tilde{\gamma}\delta t}\mathbf{p}_{n,1} + \sqrt{\frac{1}{\beta}(1-e^{-2\tilde{\gamma}\delta t})}\mathbf{M}^{1/2}\boldsymbol{\mu}_{n+1} \\
\mathbf{x}_{n+1} &= \mathbf{x}_{n,1} + \frac{\delta t}{2}\mathbf{M}^{-1}\mathbf{p}_{n+1}
\end{aligned} \qquad (8)$$

with the order $B(\delta t) A\left(\frac{1}{2}\delta t\right) O(\delta t) A\left(\frac{1}{2}\delta t\right)$ in a time step. Here $\boldsymbol{\mu}_{n+1}$ is a standard normal random vector. The random vector in each time step is independent. Arranging Eq. (8) into a compact expression leads to

$$\begin{aligned}
\mathbf{x}_{n+1} &= \mathbf{x}_n + \frac{\delta t}{2}\mathbf{M}^{-1}(1+e^{-\tilde{\gamma}\delta t})[\mathbf{p}_n - \delta t \nabla U(\mathbf{x}_n)] + \frac{\delta t}{2}\sqrt{\frac{1}{\beta}(1-e^{-2\tilde{\gamma}\delta t})}\mathbf{M}^{-1/2}\boldsymbol{\mu}_{n+1} \\
\mathbf{p}_{n+1} &= e^{-\tilde{\gamma}\delta t}[\mathbf{p}_n - \delta t \nabla U(\mathbf{x}_n)] + \sqrt{\frac{1}{\beta}(1-e^{-2\tilde{\gamma}\delta t})}\mathbf{M}^{1/2}\boldsymbol{\mu}_{n+1}
\end{aligned} \qquad (9)$$

In choosing the overdamped limit $\tilde{\gamma} = +\infty$, the momentum turns to be a random vector from the Maxwell distribution, i.e., $\mathbf{p}_{n+1} = \sqrt{\frac{1}{\beta}}\mathbf{M}^{1/2}\boldsymbol{\mu}_{n+1}$. Then the update of the configurational variable in a time step becomes

$$\mathbf{x}_{n+1} = \mathbf{x}_n - \frac{\delta t^2}{2}\mathbf{M}^{-1}\nabla U(\mathbf{x}_n) + \frac{\delta t}{2}\sqrt{\frac{1}{\beta}}\mathbf{M}^{-1/2}(\boldsymbol{\mu}_n + \boldsymbol{\mu}_{n+1}) \ . \qquad (10)$$

Note that the friction coefficient and time interval of overdamped Langevin equation are $\gamma$ and $\Delta t$, respectively. If we take the time interval of underdamped Langevin dynamics as $\delta t = \sqrt{\frac{2\Delta t}{\gamma}}$, Eq. (10) converts into

$$\mathbf{x}_{n+1} = \mathbf{x}_n - \frac{\Delta t}{\gamma}\mathbf{M}^{-1}\nabla U(\mathbf{x}_n) + \sqrt{\frac{\Delta t}{2\beta\gamma}}\mathbf{M}^{-1/2}(\mathbf{\mu}_n + \mathbf{\mu}_{n+1}),  \qquad (11)$$

which is a numerical solution of overdamped Langevin equation in a time interval $\Delta t$. For convenience, we call this algorithm "BAOA-limit". Remark that BAOA-limit is non-Markovian owing to the correlation between the stochastic part $\mathbf{\mu}_n + \mathbf{\mu}_{n+1}$ in a couple of time steps. This algorithm was first proposed in Ref. 13 by the overdamped limit of BAOAB, and rederived from the postprocessed integrators in Ref. 15. It was further studied from the perspective of the high accuracy for the stationary distribution[8, 14, 17, 23]. Our analysis here clearly shows the equivalence of BAOA with BAOAB in the configurational space.

Consider the Euler-Maruyama method (EM), which is the most common numerical algorithm for overdamped Langevin dynamics. The update of the configurational variable in a time step reads

$$\mathbf{x}_{n+1} = \mathbf{x}_n - \frac{\Delta t}{\gamma}\mathbf{M}^{-1}\nabla U(\mathbf{x}_n) + \sqrt{\frac{2\Delta t}{\beta\gamma}}\mathbf{M}^{-1/2}\mathbf{\mu}_n. \qquad (12)$$

Using the analysis similar to that of BAOA-limit, one can verify that EM is actually OBAB-limit, also in choosing $\tilde{\gamma} = +\infty$ and $\delta t = \sqrt{\frac{2\Delta t}{\gamma}}$. In the same overdamped limit condition, OABA-limit algorithm has also been introduced in Ref. 16. OABA-limit takes the form of

$$\mathbf{x}_{n+1} = \mathbf{x}_n - \frac{\Delta t}{\gamma}\mathbf{M}^{-1}\nabla U\left(\mathbf{x}_n + \sqrt{\frac{\Delta t}{2\beta\gamma}}\mathbf{M}^{-1/2}\mathbf{\mu}_n\right) + \sqrt{\frac{2\Delta t}{\beta\gamma}}\mathbf{M}^{-1/2}\mathbf{\mu}_n. \qquad (13)$$

EM and OABA-limit are Markovian.

In the rest of this paper, we compare the long-time behaviour of BAOA-limit [Eq. (11)], EM [Eq. (12)] and OABA-limit [Eq. (13)] in the harmonic system. The numerical stationary distributions of these three algorithms are explicitly derived, and the stable conditions for $\Delta t$ are verified.

## 3. Numerical stationary distributions in one-dimensional harmonic system

Exact solutions for the probability distribution of stochastic numerical algorithm are rare. Harmonic system is one of the exactly solvable models in statistical physics. In the one-dimensional harmonic system $U(x) = \frac{1}{2}m\omega^2 x^2$ where $\omega$ is the frequency, the force is simply linear $-\nabla U(x) = -m\omega^2 x$. The overdamped Langevin equation Eq. (1) for this system is now

$$\frac{dx}{dt} = -\frac{\omega^2}{\gamma}x^2 + \sqrt{\frac{2}{\beta\gamma m}}\eta(t) . \tag{14}$$

The Fokker-Planck equation Eq. (4) becomes

$$\frac{\partial}{\partial t}\rho(x,t) = \frac{\partial}{\partial x}\left[\frac{\omega^2}{\gamma}x \cdot \rho(x,t) + \frac{1}{\beta\gamma m}\frac{\partial \rho(x,t)}{\partial x}\right] . \tag{15}$$

Both Eq. (14) and Eq. (15) can be exactly solved. We treat $\eta(t)$ as a "function" of $t$, then the explicit expression of the solution for Eq. (14) can be obtained

$$x(t) = e^{-\frac{\omega^2 t}{\gamma}}x_0 + \int_0^t e^{-\frac{\omega^2(t-s)}{\gamma}}\sqrt{\frac{2}{\beta\gamma m}}\eta(s)ds \tag{16}$$

with the initial condition $x_0$. The Wiener process guarantees that $\int_0^t e^{-\frac{\omega^2(t-s)}{\gamma}}\sqrt{\frac{2}{\beta\gamma m}}\eta(s)ds$ is a normal random variable. Therefore, the probability distribution of $x(t)$ is the Gaussian distribution with the mean $e^{-\frac{\omega^2 t}{\gamma}}x_0$ and the variance

$$\left\langle\left[\int_0^t e^{-\frac{\omega^2(t-s)}{\gamma}}\sqrt{\frac{2}{\beta\gamma m}}\eta(s)ds\right]^2\right\rangle = \frac{2}{\beta\gamma m}\int_0^t\int_0^t e^{-\frac{\omega^2(t-s)}{\gamma}}e^{-\frac{\omega^2(t-s')}{\gamma}}\langle\eta(s)\eta(s')\rangle dsds'$$

$$= \frac{2}{\beta\gamma m}\int_0^t e^{-\frac{2\omega^2(t-s)}{\gamma}}ds \tag{17}$$

$$= \frac{1}{\beta m\omega^2}\left(1 - e^{-\frac{2\omega^2 t}{\gamma}}\right).$$

The time-dependent probability density is then

$$\rho(x,t) = \sqrt{\frac{\beta m\omega^2}{2\pi\left(1-e^{-\frac{2\omega^2 t}{\gamma}}\right)}} \exp\left\{-\beta \frac{1}{2} m\omega^2 \frac{1}{1-e^{-\frac{2\omega^2 t}{\gamma}}}\left[x - e^{-\frac{\omega^2 t}{\gamma}} x_0\right]^2\right\}. \quad (18)$$

One can easily examine that $\rho(x,t)$ in Eq. (18) is the solution of the Fokker-Planck equation Eq. (15) with the initial condition $\rho(x,0) = \delta(x-x_0)$. As expected, the long-time limit $t \to \infty$ of $\rho(x,t)$ leads to the Boltzmann distribution

$$\rho_{st}(x) = \sqrt{\frac{\beta m\omega^2}{2\pi}} \exp\left[-\beta \frac{1}{2} m\omega^2 x^2\right]. \quad (19)$$

Now we turn to the stationary distributions for the numerical algorithms with a finite time interval $\Delta t$. The discrete time evolution of BAOA-limit algorithm in a time interval [Eq. (11)] is now

$$x_{n+1} = \left(1 - \frac{\omega^2 \Delta t}{\gamma}\right) x_n + \sqrt{\frac{\Delta t}{2\beta\gamma m}} (\mu_n + \mu_{n+1}). \quad (20)$$

The discrete time trajectory begins from the initial condition $x_0$, then it is straightforward to show

$$x_n = \left(1 - \frac{\omega^2 \Delta t}{\gamma}\right)^n x_0 + \sqrt{\frac{\Delta t}{2\beta\gamma m}} \sum_{j=0}^{n-1} \left(1 - \frac{\omega^2 \Delta t}{\gamma}\right)^{n-j-1} (\mu_j + \mu_{j+1}). \quad (21)$$

Rearranging Eq. (21) produces

$$\begin{aligned} x_n &= \left(1 - \frac{\omega^2 \Delta t}{\gamma}\right)^n x_0 \\ &+ \sqrt{\frac{\Delta t}{2\beta\gamma m}} \left\{\left(1 - \frac{\omega^2 \Delta t}{\gamma}\right)^{n-1} \mu_0 + \sum_{j=1}^{n-1}\left[\left(1 - \frac{\omega^2 \Delta t}{\gamma}\right)^{n-j-1} + \left(1 - \frac{\omega^2 \Delta t}{\gamma}\right)^{n-j}\right]\mu_j + \mu_n\right\}. \end{aligned} \quad (22)$$

The configurational space point $x_n$ at every step of the trajectory is a linear combination of standard normal random variables $\mu_j$, thus $x_n$ satisfies the Gaussian distribution. Denote this probability distribution as $\rho_n(x)$. Since $\rho_n(x)$ is a Gaussian distribution, it is determined directly by the mean $\langle x_n \rangle$ and the variance $\langle (x_n - \langle x_n \rangle)^2 \rangle$. From Eq. (22) it is easy to obtain the mean and the variance

$$\langle x_n \rangle = \left(1 - \frac{\omega^2 \Delta t}{\gamma}\right)^n x_0 ,  \qquad (23)$$

$$\langle (x_n - \langle x_n \rangle)^2 \rangle = \frac{\Delta t}{2\beta\gamma m} \left\{ \left(1 - \frac{\omega^2 \Delta t}{\gamma}\right)^{2n-2} + \sum_{j=1}^{n-1} \left[ \left(1 - \frac{\omega^2 \Delta t}{\gamma}\right)^{n-j-1} + \left(1 - \frac{\omega^2 \Delta t}{\gamma}\right)^{n-j} \right]^2 + 1 \right\}$$

$$= \frac{\Delta t}{2\beta\gamma m} \left[ \left(1 - \frac{\omega^2 \Delta t}{\gamma}\right)^{2n-2} + \left(2 - \frac{\omega^2 \Delta t}{\gamma}\right)^2 \sum_{j=1}^{n-1} \left(1 - \frac{\omega^2 \Delta t}{\gamma}\right)^{2n-2j-2} + 1 \right] \qquad (24)$$

$$= \frac{1}{\beta m \omega^2} \left[ 1 - \left(1 - \frac{\omega^2 \Delta t}{\gamma}\right)^{2n-1} \right].$$

Taking the long-time limit $n \to \infty$ we expect $\rho_n(x)$ will approach to the stationary distribution of BAOA-limit. It is obvious that the stationary distribution exists when the stable condition $\left|1 - \frac{\omega^2 \Delta t}{\gamma}\right| < 1$ is satisfied, i.e.,

$$\frac{\omega^2 \Delta t}{\gamma} < 2 . \qquad (25)$$

Within this stable regime of $\Delta t$, the long-time limit of the mean and the variance is simply

$$\langle x_n \rangle \xrightarrow{n \to \infty} 0, \quad \langle (x_n - \langle x_n \rangle)^2 \rangle \xrightarrow{n \to \infty} \frac{1}{\beta m \omega^2} . \qquad (26)$$

The explicit expression for the stationary distribution is then

$$\rho_{\text{st}}^{\text{BAOA-limit}}(x) = \sqrt{\frac{\beta m \omega^2}{2\pi}} \exp\left[-\beta \frac{1}{2} m \omega^2 x^2\right] , \qquad (27)$$

which is exactly the Boltzmann distribution Eq. (19) of the system.

Similar analysis can be applied to EM and OABA-limit. The mean and the variance of $\rho_n(x)$ for EM is

$$\langle x_n \rangle = \left(1 - \frac{\omega^2 \Delta t}{\gamma}\right)^n x_0, \quad \langle (x_n - \langle x_n \rangle)^2 \rangle = \frac{1}{\beta m \omega^2} \frac{1}{1 - \frac{\omega^2 \Delta t}{2\gamma}} \left[1 - \left(1 - \frac{\omega^2 \Delta t}{\gamma}\right)^{2n}\right]. \tag{28}$$

The stationary distribution in the long-time limit within the stable regime Eq. (25) is

$$\rho_{st}^{EM}(x) = \sqrt{\frac{\beta m \omega^2 \left(1 - \frac{\omega^2 \Delta t}{2\gamma}\right)}{2\pi}} \exp\left[-\beta \frac{1}{2} m \omega^2 x^2 \left(1 - \frac{\omega^2 \Delta t}{2\gamma}\right)\right]. \tag{29}$$

The result for OABA-limit is

$$\langle x_n \rangle = \left(1 - \frac{\omega^2 \Delta t}{\gamma}\right)^n x_0, \quad \langle (x_n - \langle x_n \rangle)^2 \rangle = \frac{1}{\beta m \omega^2} \left(1 - \frac{\omega^2 \Delta t}{2\gamma}\right) \left[1 - \left(1 - \frac{\omega^2 \Delta t}{\gamma}\right)^{2n}\right] \tag{30}$$

and

$$\rho_{st}^{OABA\text{-limit}}(x) = \sqrt{\frac{\beta m \omega^2}{2\pi \left(1 - \frac{\omega^2 \Delta t}{2\gamma}\right)}} \exp\left[-\beta \frac{1}{2} m \omega^2 x^2 \frac{1}{1 - \frac{\omega^2 \Delta t}{2\gamma}}\right] \tag{31}$$

within the stable regime Eq. (25). Obviously, for both EM [Eq. (29)] and OABA-limit [Eq. (31)], the infinitesimal time interval limit $\Delta t \to 0$ leads to the exact Boltzmann distribution. Comparison of the accuracy associated with $\Delta t$ of these three algorithms is straightforward. BAOA-limit algorithm generates the exact Boltzmann distribution $\rho_{st}(x)$ even when $\Delta t$ is finite, as long as the stable condition is satisfied. EM and OABA-limit fail to do so. Numerical error of the stationary distribution owing to the finite $\Delta t$ is the first order for both EM and OABA-limit. The accuracy analysis can be summarized as

$$\begin{aligned} \rho_{st}^{BAOA\text{-limit}}(x) &= \rho_{st}(x), \\ \rho_{st}^{EM}(x) &= \rho_{st}(x) \times [1 + O(\Delta t)], \\ \rho_{st}^{OABA\text{-limit}}(x) &= \rho_{st}(x) \times [1 + O(\Delta t)]. \end{aligned} \tag{32}$$

We should remark that, Ref. 22 employed a trajectory-based approach to derive the numerical stationary distribution in the phase space, of the underdamped Langevin integrator in the harmonic system. The analysis of the discrete time overdamped Langevin trajectory in this paper can be seen as the extension of the trajectory-based approach for underdamped Langevin dynamics.

**4. Results in multi-dimensional harmonic system**

We turn to the general *k*-dimensional harmonic system $U(\mathbf{x}) = \frac{1}{2}\mathbf{x}^T\mathbf{H}\mathbf{x}$, where $\mathbf{H}$ is the symmetric positive definite Hessian matrix. The discrete time evolution in $\Delta t$ of BAOA-limit is now

$$\mathbf{x}_{n+1} = \mathbf{x}_n - \frac{\Delta t}{\gamma}\mathbf{M}^{-1}\mathbf{H}\mathbf{x}_n + \sqrt{\frac{\Delta t}{2\beta\gamma}}\mathbf{M}^{-1/2}(\boldsymbol{\mu}_n + \boldsymbol{\mu}_{n+1}) \ . \tag{33}$$

One can obtain the configurational space point $\mathbf{x}_n$, similar to the expressions in Eqs. (21) and (22) for one-dimensional case. Here we employ an alternative approach by using the normal mode coordinate transformation. The characteristic frequencies of the normal modes can be obtained from the eigenvalues of the matrix $\mathbf{M}^{-1/2}\mathbf{H}\mathbf{M}^{-1/2}$, i.e.,

$$\mathbf{M}^{-1/2}\mathbf{H}\mathbf{M}^{-1/2}\mathbf{T} = \mathbf{T}[\boldsymbol{\omega}^2] \ . \tag{34}$$

Here $\mathbf{T}$ is an orthogonal matrix consisting of the eigenvectors of $\mathbf{M}^{-1/2}\mathbf{H}\mathbf{M}^{-1/2}$, and

$$[\boldsymbol{\omega}^2] = \begin{bmatrix} \omega_1^2 & & \\ & \ddots & \\ & & \omega_k^2 \end{bmatrix} \tag{35}$$

is a diagonal matrix consisting of the eigenvalues with the characteristic frequency $\omega_i$ for each normal mode. Then the normal mode coordinate transformation is

$$\mathbf{q} = \mathbf{T}^T\mathbf{M}^{1/2}\mathbf{x} \tag{36}$$

and the Hessian matrix $\mathbf{H}$ satisfies

$$\mathbf{H} = \mathbf{M}^{1/2}\mathbf{T}\left[\boldsymbol{\omega}^2\right]\mathbf{T}^T\mathbf{M}^{1/2} . \tag{37}$$

Rewrite the discrete time evolution Eq. (33) in the normal mode configurational space

$$\mathbf{q}_{n+1} = \mathbf{q}_n - \frac{\Delta t}{\gamma}\left[\boldsymbol{\omega}^2\right]\mathbf{q}_n + \sqrt{\frac{\Delta t}{2\beta\gamma}}\mathbf{T}^T\left(\boldsymbol{\mu}_n + \boldsymbol{\mu}_{n+1}\right) . \tag{38}$$

Denote $\tilde{\boldsymbol{\mu}}_n = \mathbf{T}^T\boldsymbol{\mu}_n$ and we find $\tilde{\boldsymbol{\mu}}_n$ is still a standard normal random vector. It is easy to verify that, the evolution of each degree of normal mode coordinate is independent in Eq. (38). For the degree $i$, the evolution reads

$$q_{i,n+1} = \left(1 - \frac{\omega_i^2 \Delta t}{\gamma}\right)q_{i,n} + \sqrt{\frac{\Delta t}{2\beta\gamma}}\left(\tilde{\mu}_{i,n} + \tilde{\mu}_{i,n+1}\right) , \tag{39}$$

which takes the same form of Eq. (20) with a unit mass for the one-dimensional case. Then we can directly employ Eq. (27) to give the stationary distribution of $q_i$

$$\rho_{\text{st}}^{\text{BAOA-limit}}(q_i) = \sqrt{\frac{\beta\omega_i^2}{2\pi}}\exp\left[-\beta\frac{1}{2}\omega_i^2 q_i^2\right] \tag{40}$$

with the stable regime

$$\frac{\omega_i^2 \Delta t}{\gamma} < 2 . \tag{41}$$

Since the distribution of each degree of normal mode coordinate is independent, the stationary distribution in the normal mode configurational space is then

$$\rho_{\text{st}}^{\text{BAOA-limit}}(\mathbf{q}) = \left(\frac{\beta}{2\pi}\right)^{k/2}\left(\det\left[\boldsymbol{\omega}^2\right]\right)^{1/2}\exp\left\{-\beta\frac{1}{2}\mathbf{q}^T\left[\boldsymbol{\omega}^2\right]\mathbf{q}\right\} . \tag{42}$$

By the inverse coordinate transformation $\mathbf{x} = \mathbf{M}^{-1/2}\mathbf{T}\mathbf{q}$, we obtain

$$\begin{aligned}\rho_{\text{st}}^{\text{BAOA-limit}}(\mathbf{x}) &= \left(\frac{\beta}{2\pi}\right)^{k/2}\left(\det\left[\boldsymbol{\omega}^2\right]\right)^{1/2}\left(\det\mathbf{M}\right)^{1/2}\exp\left\{-\beta\frac{1}{2}\mathbf{x}^T\mathbf{M}^{1/2}\mathbf{T}\left[\boldsymbol{\omega}^2\right]\mathbf{T}^T\mathbf{M}^{1/2}\mathbf{x}\right\} \\ &= \left(\frac{\beta}{2\pi}\right)^{k/2}\left(\det\mathbf{H}\right)^{1/2}\exp\left\{-\beta\frac{1}{2}\mathbf{x}^T\mathbf{H}\mathbf{x}\right\}.\end{aligned} \tag{43}$$

The result demonstrates that BAOA-limit generates the exact Boltzmann distribution in the stationary state, when the stable condition of $\Delta t$ Eq. (41) is satisfied for each characteristic frequency $\omega_i$.

The stationary distributions of EM and OABA-limit can be derived using the similar technique

$$\rho_{st}^{EM}(\mathbf{x}) = \left(\frac{\beta}{2\pi}\right)^{k/2} \left(\det\left[\left(\mathbf{1} - \mathbf{HM}^{-1}\frac{\Delta t}{2\gamma}\right)\mathbf{H}\right]\right)^{1/2} \exp\left\{-\beta\frac{1}{2}\mathbf{x}^T\left(\mathbf{1} - \mathbf{HM}^{-1}\frac{\Delta t}{2\gamma}\right)\mathbf{Hx}\right\}, \quad (44)$$

$$\rho_{st}^{OABA\text{-limit}}(\mathbf{x}) = \left(\frac{\beta}{2\pi}\right)^{k/2} \left(\det\left[\left(\mathbf{1} - \mathbf{HM}^{-1}\frac{\Delta t}{2\gamma}\right)^{-1}\mathbf{H}\right]\right)^{1/2} \exp\left\{-\beta\frac{1}{2}\mathbf{x}^T\left(\mathbf{1} - \mathbf{HM}^{-1}\frac{\Delta t}{2\gamma}\right)^{-1}\mathbf{Hx}\right\}. \quad (45)$$

For both algorithms, the stable condition is Eq. (41) for each characteristic frequency $\omega_i$, and the infinitesimal time interval limit $\Delta t \to 0$ leads to the exact Boltzmann distribution. It is trivial to find that the results proposed here for all the algorithms are consistent with that of the one-dimensional case when $k$ reduces to 1. The advantage in the long-time behaviour of BAOA-limit algorithm is shown clearly. For both EM and OABA-limit, numerical error in the stationary distribution is the first order of $\Delta t$ in comparison to the exact Boltzmann distribution. Therefore, the accuracy of these algorithms is consistent with that of the one-dimensional case as shown in Eq. (32).

## 5. Conclusions

The high accuracy of BAOA-limit (or BAOAB-limit) algorithm in the equilibrium configurational sampling has been confirmed in previous studies. In this paper, we present the evidence for this statement in an exactly solvable system. Within the stable regime of $\Delta t$, the numerical stationary distribution generated by BAOA-limit algorithm of the harmonic system is exactly the Boltzmann distribution, both for one-dimensional and multi-dimensional cases. EM (OBAB-limit) and OABA-limit, the other two algorithms considered here for comparison, both lead to the stationary distribution with the first order error of $\Delta t$. To derive the explicit expression of the numerical stationary distribution, we employ the stochastic analysis of the discrete time trajectory. This method may be extended to other numerical solutions of stochastic differential equation.

Besides in sampling the equilibrium canonical ensemble, overdamped Langevin equation can also be employed in the nonstationary process. For example, a conditional overdamped Langevin equation for generating transition paths from an initial state to a given final state is proposed[31-35]. Nonstationary distribution of overdamped Langevin equation deserves further study. The performance of BAOA-limit algorithm in sampling the transition path ensemble is of interest in future works.

## Statements and Declarations

### Ethical Approval

Not applicable.

### Availability of supporting data

Not applicable. No new data were created or analyzed in this study.

### Competing interests

The authors have no competing interests to declare that are relevant to the content of this paper.

### Funding

This work was supported by Guangdong Basic and Applied Basic Research Foundation (Grant No. 2021A1515010328), Key-Area Research and Development Program of Guangdong

Province (Grant No. 2020B010183001), and National Natural Science Foundation of China (Grant No. 12074126).

**Authors' contributions**

De-Zhang Li finished the main text of this paper. All authors reviewed this paper.

**Acknowledgments**

De-Zhang Li thanks Dr. Yun-Feng Xiong for the private communication of the content of this paper.

## References

1. P. Langevin, "Sur la théorie du mouvement brownien," C. R. Acad. Sci. (Paris) **146**, 530-533 (1908).

2. D. S. Lemons and A. Gythiel, "Paul Langevin's 1908 paper "On the Theory of Brownian Motion" ["Sur la théorie du mouvement brownien," C. R. Acad. Sci. (Paris) 146, 530–533 (1908)]," Am. J. Phys. **65** (11), 1079-1081 (1997). https://doi.org/10.1119/1.18725

3. A. Einstein, "Über die von der molekularkinetischen Theorie der Wärme geforderte Bewegung von in ruhenden Flüssigkeiten suspendierten Teilchen," Ann. Phys. **322** (8), 549-560 (1905). https://doi.org/10.1002/andp.19053220806

4. A. Einstein, "Zur Theorie der Brownschen Bewegung," Ann. Phys. **324** (2), 371-381 (1906). https://doi.org/10.1002/andp.19063240208

5. M. v. Smoluchowski, "Zur kinetischen Theorie der Brownschen Molekularbewegung und der Suspensionen," Ann. Phys. **326** (14), 756-780 (1906). https://doi.org/10.1002/andp.19063261405

6.  N. G. v. Kampen, *Stochastic Processes in Physics and Chemistry*, 3rd ed. (Elsevier, Amsterdam, 2009).

7.  R. Zwanzig, *Nonequilibrium statistical mechanics*. (Oxford University Press, New York, 2001).

8.  B. Leimkuhler and M. Sachs, "Efficient Numerical Algorithms for the Generalized Langevin Equation," SIAM J. Sci. Comput. **44** (1), A364-A388 (2022). https://doi.org/10.1137/20M138497X

9.  A. D. Fokker, "Die mittlere Energie rotierender elektrischer Dipole im Strahlungsfeld," Ann. Phys. **348** (5), 810-820 (1914). https://doi.org/10.1002/andp.19143480507

10. V. Planck, "Über einen Satz der statistischen Dynamik und seine Erweiterung in der Quantentheorie," Sitzungsber. Preuss. Akad. Wiss. **24**, 324-341 (1917).

11. H. Risken, *The Fokker-Planck Equation: Methods of Solution and Applications*. (Springer-Verlag, Berlin, 1989).

12. G. A. Pavliotis, *Stochastic Processes and Applications: Diffusion Processes, the Fokker-Planck and Langevin Equations*. (Springer, New York, 2014).

13. B. Leimkuhler and C. Matthews, "Rational Construction of Stochastic Numerical Methods for Molecular Sampling," Appl. Math. Res. Express **2013** (1), 34-56 (2013). https://doi.org/10.1093/amrx/abs010

14. B. Leimkuhler, C. Matthews and M. V. Tretyakov, "On the long-time integration of stochastic gradient systems," Proc. R. Soc. A **470** (2170), 20140120 (2014). https://doi.org/10.1098/rspa.2014.0120

15. G. Vilmart, "Postprocessed Integrators for the High Order Integration of Ergodic SDEs," SIAM J. Sci. Comput. **37** (1), A201-A220 (2015). https://doi.org/10.1137/140974328

16. M. Fathi and G. Stoltz, "Improving dynamical properties of metropolized discretizations of overdamped Langevin dynamics," Numer. Math. **136** (2), 545-602 (2017). https://doi.org/10.1007/s00211-016-0849-3

17. X. Shang and M. Kröger, "Time Correlation Functions of Equilibrium and Nonequilibrium Langevin Dynamics: Derivations and Numerics Using Random Numbers," SIAM Rev. **62** (4), 901-935 (2020). https://doi.org/10.1137/19M1255471

18. G. E. Uhlenbeck and L. S. Ornstein, "On the Theory of the Brownian Motion," Phys. Rev. **36** (5), 823-841 (1930). https://doi.org/10.1103/PhysRev.36.823

19. M. C. Wang and G. E. Uhlenbeck, "On the Theory of the Brownian Motion II," Rev. Mod. Phys. **17** (2-3), 323-342 (1945). https://doi.org/10.1103/RevModPhys.17.323

20. L. Verlet, "Computer "experiments" on classical fluids. I. Thermodynamical properties of Lennard-Jones molecules," Phys. Rev. **159** (1), 98 (1967). https://doi.org/10.1103/PhysRev.159.98

21. B. Leimkuhler and C. Matthews, "Robust and efficient configurational molecular sampling via Langevin dynamics," J. Chem. Phys. **138** (17), 174102 (2013). https://doi.org/10.1063/1.4802990

22. D. Li, X. Han, Y. Chai, C. Wang, Z. Zhang, Z. Chen, J. Liu and J. Shao, "Stationary state distribution and efficiency analysis of the Langevin equation via real or virtual dynamics," J. Chem. Phys. **147** (18), 184104 (2017). https://doi.org/10.1063/1.4996204

23. B. Leimkuhler, C. Matthews and G. Stoltz, "The computation of averages from equilibrium and nonequilibrium Langevin molecular dynamics," IMA J. Numer. Anal. **36** (1), 13-79 (2016). https://doi.org/10.1093/imanum/dru056

24. N. Grønbech-Jensen and O. Farago, "A simple and effective Verlet-type algorithm for simulating Langevin dynamics," Mol. Phys. **111** (8), 983-991 (2013). https://doi.org/10.1080/00268976.2012.760055

25. J. Liu, D. Li and X. Liu, "A simple and accurate algorithm for path integral molecular dynamics with the Langevin thermostat," J. Chem. Phys. **145** (2), 024103 (2016). https://doi.org/10.1063/1.4954990

26. Z. Zhang, X. Liu, Z. Chen, H. Zheng, K. Yan and J. Liu, "A unified thermostat scheme for efficient configurational sampling for classical/quantum canonical ensembles via molecular dynamics," J. Chem. Phys. **147** (3), 034109 (2017). https://doi.org/10.1063/1.4991621

27. D. Li, Z. Chen, Z. Zhang and J. Liu, "Understanding Molecular Dynamics with Stochastic Processes via Real or Virtual Dynamics," Chin. J. Chem. Phys. **30** (6), 735-760 (2017). https://doi.org/10.1063/1674-0068/30/cjcp1711223

28. N. Grønbech-Jensen, "Complete set of stochastic Verlet-type thermostats for correct Langevin simulations," Mol. Phys. **118** (8), e1662506 (2020). https://doi.org/10.1080/00268976.2019.1662506

29. Z. Zhang, K. Yan, X. Liu and J. Liu, "A leap-frog algorithm-based efficient unified thermostat scheme for molecular dynamics," Chin. Sci. Bull. **63** (33), 3467-3483 (2018). https://doi.org/10.1360/N972018-00908

30. Z. Zhang, X. Liu, K. Yan, M. E. Tuckerman and J. Liu, "Unified Efficient Thermostat Scheme for the Canonical Ensemble with Holonomic or Isokinetic Constraints via Molecular Dynamics," J. Phys. Chem. A **123** (28), 6056-6079 (2019). https://doi.org/10.1021/acs.jpca.9b02771

31. H. Orland, "Generating transition paths by Langevin bridges," J. Chem. Phys. **134** (17), 174114 (2011). https://doi.org/10.1063/1.3586036

32. S. N. Majumdar and H. Orland, "Effective Langevin equations for constrained stochastic processes," J. Stat. Mech. **2015** (6), P06039 (2015). https://doi.org/10.1088/1742-5468/2015/06/p06039

33. M. Delarue, P. Koehl and H. Orland, "Ab initio sampling of transition paths by conditioned Langevin dynamics," J. Chem. Phys. **147** (15), 152703 (2017). https://doi.org/10.1063/1.4985651

34. R. Elber, D. E. Makarov and H. Orland, *Molecular Kinetics in Condensed Phases: Theory, Simulation, and Analysis*. (Wiley, Hoboken, NJ, 2020).

35. P. Koehl and H. Orland, "Sampling constrained stochastic trajectories using Brownian bridges," J. Chem. Phys. **157** (5), 054105 (2022). https://doi.org/10.1063/5.0102295